\documentclass[sigplan,nonacm]{acmart}

\usepackage[utf8]{inputenc}
\usepackage{url}
\usepackage{geometry}
\usepackage{listings}
\usepackage[normalem]{ulem}
\usepackage{siunitx}            
\usepackage{courier}            
\usepackage{comment}
\usepackage{helvet} 
\usepackage{xspace}

\usepackage{tikz}
\usepackage{amsmath}

\usepackage{subcaption}
\usepackage{xspace}
\usepackage{cleveref}
\usepackage{multirow}

\usepackage{lineno}
\usepackage{hyperref}
\usepackage{float}
\usepackage{svg}
\usepackage{graphicx}
\usepackage{color}
\usepackage{siunitx}
\usepackage{xparse}
\usepackage{enumitem}

\newcommand{\eg}{e.g.,\xspace}
\newcommand{\ie}{i.e.,\xspace}

\def\sys{TVA\xspace}

\def\numcases{2}

\newcommand{\grumbler}[3]{}

  {\begin{enumerate}[itemsep=-0pt, parsep=0pt, topsep=0pt, leftmargin=2pc]}
  {\end{enumerate}}

\newenvironment{smitemize}%
  {\begin{list}{$\bullet$}%
    {\setlength{\parsep}{0pt}%
      \setlength{\topsep}{0pt}%
      \setlength{\leftmargin}{2pc}%
      \setlength{\itemsep}{1pt}}}
  {\end{list}}

\usepackage{pslatex}

\begin{abstract}
Rewriting x86\_64 binaries---whether for security hardening, dynamic instrumentation, or performance profiling---is notoriously difficult due to variable-length instructions, interleaved code and data, and indirect jumps to arbitrary byte offsets. Existing solutions (e.g., ``superset disassembly'') ensure soundness but incur significant overhead and produce large rewritten binaries, especially for on-the-fly instrumentation. 
This paper addresses these challenges by introducing the \emph{Time Variance Authority}~(TVA), which leverages Intel's Control-Flow Enforcement Technology~(CET). By recognizing \texttt{endbr64} as the only valid indirect jump target, TVA prunes spurious disassembly paths while preserving soundness and emulates CET constraints on processors lacking native CET support, effectively mitigating ROP/JOP exploits without new hardware. 
We implement TVA by modernizing the Multiverse rewriter for 64-bit Linux. Our evaluation on SPEC~CPU2017 and real-world applications shows that TVA-guided rewriting achieves up to 1.3$\times$ faster instrumentation time. These results underscore TVA's feasibility as a high-performance, uprobes-free alternative for robust x86\_64 binary analysis and rewriting.
\end{abstract}
\author{Brian Zhao, Yiwei Yang, Yusheng Zheng, Andi Quinn}
\affiliation{UC Santa Cruz \country{USA}}

\begin{document}


\title{Exploiting Control-flow Enforcement Technology for Sound and Precise Static Binary Disassembly}
\maketitle

\section{Introduction}

Modern software ecosystems routinely rely on post-compilation code
modifications for a variety of purposes: adding security hardening to
prevent advanced exploits, injecting instrumentation for fine-grained
performance profiling, or supporting fault isolation in closed-source
environments. With the explosive growth of third-party libraries and
precompiled modules, it is no longer feasible to assume source-level
access for every component. As a result, \emph{binary rewriting} has
emerged as a critical mechanism to retrofit protections and analyses
into \textit{commercial-off-the-shelf} (COTS) software, yet performing it safely
and efficiently on x86\_64 architectures remains an open challenge.

Rewriting x86\_64 binaries for security hardening, instrumentation, or
fault isolation poses significant difficulties when source code is
unavailable. As shown in \Cref{tab:related-work}, many static tools
(e.g., RedFat~\cite{duck2022hardening}, E9AFL~\cite{e9afl}) apply
post-compilation transformations to detect memory errors or enable
coverage-guided fuzzing, while others (NaCL~\cite{nacl},
WebAssembly~\cite{webassembly}) rely on specialized compiler toolchains
to ensure sandboxing. Yet, \emph{soundness}---fully capturing all valid
execution paths of an arbitrary binary---remains notoriously difficult.
The primary obstacle is that x86\_64 permits variable-length
instructions, interleaving of code and data, and indirect jumps to
practically any byte offset. Such complexities typically force rewriting
frameworks to over-approximate the set of possible instruction
boundaries, leading to bloated analysis times and excessive runtime
overhead.

An extreme example of over-approximation is \emph{superset disassembly},
as implemented in Multiverse~\cite{bauman2018superset}, which ensures
that any valid offset an indirect jump \emph{could} target is treated as
real code. Although this guarantees correctness, the resulting
instrumentation often suffers from high compilation overhead and
spurious execution paths. Dynamic binary instrumentation systems
such as Intel Pin~\cite{pin} and DynamoRIO~\cite{dynamorio} also
address soundness by intercepting instructions at runtime, but may
incur substantial performance penalties.

Recent hardware-based Control-Flow Integrity (CFI) mechanisms offer a
way to limit this over-approximation. Intel's Control-Flow Enforcement
Technology~(CET)~\cite{cet,kim2022d,kim2025towards} uses \texttt{endbr64}
instructions to denote valid \emph{indirect} jump destinations, thereby
making spurious transfers (e.g., jumping into the middle of an
instruction) illegal in hardware. In principle, if a CET-enabled binary
is rewritten, the rewriter can prune offsets that do not begin with
\texttt{endbr64}. Moreover, shadow stacks~\cite{cet-analysis}---whether
in hardware or software---further mitigate return-oriented programming
(ROP) by ensuring that return addresses match their corresponding call
sites.

In this paper, we present the \emph{Time Variance Authority (TVA)}, a
CET-driven rewriting framework that builds on Multiverse and applies to
both static and dynamic instrumentation scenarios. By treating
\texttt{endbr64} as the only legitimate targets of indirect jumps, TVA
avoids the worst-case explosion in disassembly coverage while retaining
fully sound rewriting. Additionally, TVA enables software emulation of
CET constraints for processors lacking native CET support; in so doing,
it enforces shadow stack checks and indirect-branch validity without
requiring special hardware extensions. We implement our system as a
drop-in enhancement over existing superset-disassembly workflows,
leveraging Multiverse's robust approach for code layout while plugging
in CET logic to reduce unnecessary paths. Our code is available at \url{https://github.com/SlugLab/tva}.

\begin{quote}
    \textit{TVA is depicted as a group of timeline monitors tasked with preventing the existence of certain timelines that are deemed too dangerous to the Multiverse.\quad \quad\quad \quad\quad \quad\quad \quad\quad --\quad Marvell}
\end{quote}

To summary our contributions:
\begin{itemize}
  \item \textbf{CET-guided precise disassembly.}
    TVA leverages CET's \texttt{endbr64} instructions as unique markers of valid indirect jump targets, eliminating unnecessary disassembly paths and preserving soundness—a key improvement over traditional superset methods.

  \item \textbf{Software CET emulation.}
    TVA enforces CET constraints purely in software, enabling robust protection against control-flow attacks (ROP/JOP) even on legacy CPUs without native CET support.

  \item \textbf{High-performance instrumentation.}
    Evaluation on SPEC CPU2017 demonstrates TVA's practical benefits, showing up to 1.3$\times$ faster instrumentation.
\end{itemize}

The rest of the paper is organized as follows. We first discuss
background and technical challenges (\Cref{sec:background}), then present
\sys's design (\Cref{sec:design}) and implementation details
(\Cref{sec:implementation}). We evaluate our system in
\Cref{sec:evaluation} and conclude in \Cref{sec:conclusion}.

\begin{table*}[ht]
\centering
\caption{Comparison of Related Binary Rewriting and Hardening Tools}
\label{tab:related-work}
\begin{tabular}{p{2cm} p{2cm} p{2cm} p{3.2cm} p{4.5cm}}
\hline
\textbf{Tool} & \textbf{Approach} & \textbf{Application} & \textbf{Key Strength} & \textbf{Limitation} \\
\hline
\hline
\textbf{RedFat} 
 & Binary rewriting of pointers  
 & Memory error detection (buffer overflows, UAF) 
 & Detects pointer-related errors at runtime 
 & Potential performance overhead due to pointer “fatness” \\
\textbf{SelectiveTaint}
 & Static binary rewriting 
 & Dynamic taint analysis 
 & Automated injection of taint labels into rewritten code 
 & Can incur high performance overhead on large, complex binaries \\
\textbf{E9AFL}
 & Static binary rewriting 
 & Fuzzing without recompilation 
 & Eliminates need for source code; direct instrumentation of binaries 
 & May not achieve complete coverage on highly obfuscated binaries \\
\textbf{NaCl}
 & Custom compiler toolchain 
 & Fault isolation / sandboxing 
 & Ensures untrusted x86 code complies with sandbox constraints 
 & Not purely binary-level rewriting; requires specialized compilation \\
\textbf{WebAssembly}
 & Custom compiler and VM environment 
 & Safe, portable execution 
 & High portability across platforms; reduced attack surface 
 & Requires re-targeting / compilation to Wasm; not a post-compilation tool \\
\hline
\end{tabular}
\end{table*}
 
\section{Background} \label{sec:background}

This section begins with a
discussion of why x86\_64 machine code complicates static analysis, followed by an overview of ROP and Intel CET. We then survey existing disassembly and rewriting 
techniques, highlighting how hardware-backed 
control-flow integrity (CFI) can prune impossible paths. Finally, we present 
key insights from Multiverse---a superset disassembly framework---and 
explain how our system builds on and refines those ideas.
\subsection{Complexities of x86\_64 Machine Code} \label{sec:x86machine}

Modern x86\_64 processors implement a Complex Instruction Set Computing (CISC) architecture. A defining feature of this architecture is the use of variable-length instructions, which range from as short as one byte (e.g., \texttt{nop}) to as long as fifteen bytes (e.g., instructions containing multiple prefixes, ModR/M bytes, and displacement/immediate fields). Unlike certain Reduced Instruction Set Computing (RISC) designs such as ARM (in certain modes), x86\_64 does not require instructions to be aligned to word or halfword boundaries; instead, an instruction may begin at any arbitrary byte offset. This unaligned nature contributes significantly to the complexity faced by static disassembly and rewriting tools.

The lack of enforced alignment allows the instruction encoding to be highly flexible and compact. However, this flexibility introduces considerable difficulties for static analysis tasks, including binary instrumentation, rewriting, and security analysis. Because each instruction is decoded sequentially, any decoding error, however slight, can propagate throughout subsequent instructions—a phenomenon known as \textit{disassembly drift}. Such drift results when the disassembler mistakenly interprets data bytes as instructions or misses genuine instructions entirely. Additionally, legitimate binary code often interleaves literal data (such as jump tables or constant pools) within instruction streams, making the differentiation between code and data challenging based solely on byte inspection.

Moreover, the variable-length and non-aligned nature of x86\_64 instructions complicates binary patching and rewriting. Any naive insertion or deletion of code bytes can inadvertently shift the position of subsequent instructions. Such shifts risk corrupting instruction sequences, overwriting essential data literals, or misaligning indirect jump and call targets. Consequently, tools performing static rewriting must incorporate advanced handling mechanisms or heuristics—such as those employed in \emph{superset disassembly}~\cite{bauman2018superset}—to maintain correctness and stability, especially when processing complex control-flow structures or position-independent code (PIC/PIE).

Figure~\ref{fig:showcase} illustrates how instructions of varying lengths are tightly packed within a binary, highlighting the necessity for accurate boundary detection during disassembly. Existing approaches like superset disassembly attempt to ensure correctness by analyzing every potential instruction offset, inevitably leading to inflated runtime overhead due to excessive coverage.

Given these challenges, a precise understanding of the underlying Instruction Set Architecture (ISA) becomes critical. The ISA defines every aspect of an instruction's representation and semantics, including opcode definitions, operand types (registers, immediate values, or memory references), and encoding formats (prefixes, ModR/M bytes, displacements, and immediate fields). Accurate disassembly and rewriting depend on reliably interpreting these ISA-specified instruction details. Misinterpretation or incorrect handling at this foundational level can jeopardize both security and correctness, emphasizing why robust ISA-aware disassembly techniques are essential for modern static binary rewriting frameworks.

In the sections that follow, we explore how emerging control-flow enforcement technologies like Intel’s Control-Flow Enforcement Technology (CET) influence static binary rewriting and disassembly processes, particularly by leveraging explicit ISA-level markers to reduce ambiguity and complexity during analysis.

\subsection{ROP attacks and CET}
\begin{figure*}[h!]
\centering
\includegraphics[width=2\columnwidth]{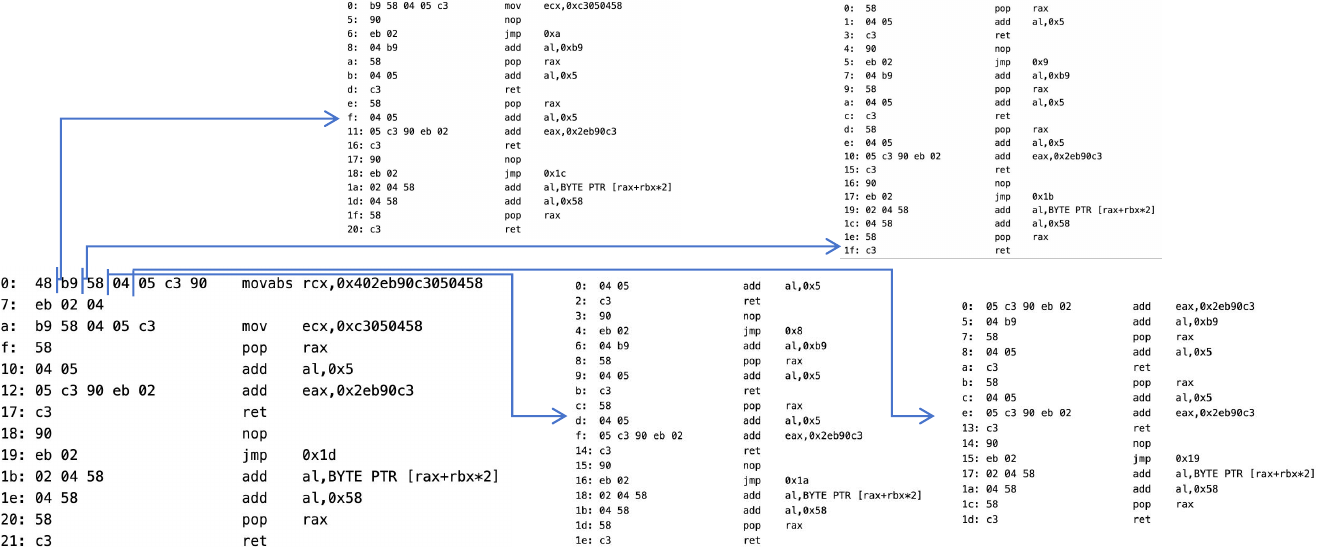}
\caption{A gadget that can be parsed from arbitrary offset.}
\label{fig:showcase}
\end{figure*}

Return-Oriented Programming (ROP) is a common exploit technique that allows an attacker 
to stitch together small code fragments (known as “gadgets”) already present in a binary, 
bypassing traditional non-executable memory protections such as DEP (Data Execution Prevention).
Figure~\ref{fig:showcase} shows a toy example of ROP payload steps in x86-64 assembly.

To mitigate such attacks, Intel introduced Control-Flow Enforcement Technology (CET). CET 
provides hardware-based control-flow integrity by imposing restrictions on indirect branch 
targets using \texttt{endbr64} instructions. Below is a simplified example showing how inserting 
\texttt{endbr64} at valid jump targets can prevent unwanted transfers of control.  \texttt{endbr64} can also acts as a \emph{hardware label} for valid
indirect jump targets. Even though CET instructions are interpreted as
no-ops on older CPUs, they provide strong hints for static tools on where
legitimate indirect jumps may land. Consequently, a static rewriting
framework can ignore offsets that lack an \texttt{endbr64} prefix,
significantly pruning disassembly paths.

\subsection{Disassembly and Rewriting Techniques}
\label{sec:disassembly}

Broadly, existing methods for rewriting x86\_64 binaries fall into two
categories:

\subsubsection{Reassemblable Disassembly}
Tools like \textbf{Uroboros}~\cite{wang2016uroboros},
\textbf{Retrowrite}~\cite{dinesh2020retrowrite},
\textbf{DDisasm}~\cite{flores2020datalog}, and
\textbf{Multiverse}~\cite{bauman2018superset} translate raw machine code
into reassemblable assembly, modify it, and then reconstruct the
binary. This can preserve much of the binary’s original structure, but
the rewriting must handle position-independent code (PIC/PIE), relocation
tables, and precise instruction boundaries.  

\subsubsection{Direct Rewriting / Instruction Punning}
Approaches such as \textbf{Zipr}~\cite{hawkins2017zipr},
\textbf{Egalito}~\cite{williams2020egalito}, \textbf{zpoline}\cite{zpoline}, \textbf{frida}\cite{frida} or \textbf{e9patch}~\cite{duck2020binary}
lift code into an intermediate representation (IR) or create out-of-line
trampolines. This strategy can reduce disruption to the original binary’s
layout but still relies on identifying correct disassembly boundaries for
jump insertion. Moreover, advanced transformations can be limited if 
the original code has deeply interwoven instructions and data.  

\subsection{Hardware-Accelerated CFI Mechanisms}
\label{sec:cfihardware}

Beyond CET, hardware support for CFI has expanded in recent CPU generations.
For example, AMD Zen3 introduces a hardware \emph{shadow stack} to validate
return addresses~\cite{ssp}. In principle, these features drastically reduce
the \emph{runtime} attack surface, as only labeled entry points are valid
targets. However, to exploit these mechanisms fully at \emph{static rewrite
time}, one must ensure that the rewritten binary respects or emulates
CET-style invariants. That is exactly what our approach does: even on CPUs
lacking CET, \texttt{endbr64} instructions can still serve as a
filter for feasible indirect jumps, and a software-maintained shadow stack
can mimic hardware protections against ROP.

\subsection{Challenges and Insights from Multiverse}
\label{sec:multiverse}

\textbf{Multiverse}~\cite{bauman2018superset} pioneered the idea of 
\emph{superset disassembly}, disassembling \emph{every} potential offset 
to guarantee no valid code path is missed. This yields a \emph{sound} 
rewrite---any legitimate indirect jump target is included---but at the cost
of major over-approximation. Indeed, many “instructions” discovered in this
manner are simply data bytes or offsets that are never actually taken at 
runtime.

Multiverse also tackled five specific challenges:
\begin{enumerate}[label=\textbf{C\arabic*}]
    \item \textbf{Static Memory Address Relocation.}
          Ensuring correct relocation for instructions referencing data
          or global symbols when the code layout changes.
    \item \textbf{Dynamic Address Computation.}
          Handling register-based address arithmetic that complicates 
          static determination of control-flow edges.
    \item \textbf{Code vs. Data Differentiation.}
          Dealing with arbitrary interleaving of code and data in 
          x86\_64 segments.
    \item \textbf{Function Pointer Handling.}
          Maintaining correct references to function pointers that may 
          cross module boundaries or exist in data sections.
    \item \textbf{Position-Independent Code (PIC/PIE).}
          Accounting for modern compilers and operating systems that 
          routinely use relative addressing and dynamic relocation.
\end{enumerate}

To address these issues, Multiverse preserves most of the original layout
(\textbf{S1}) and records an old-to-new offset mapping (\textbf{S2}), allowing
instructions or data references to be correctly redirected. For PIC/PIE
(\textbf{S5}), it defers to relative addressing so the resulting code remains
relocatable.

\paragraph{Our Key Insight.} Instead of continuing to disassemble \emph{all}
offsets (as in \textbf{S3}), we observe that many modern binaries are compiled
with \texttt{endbr64} instructions at \emph{legitimate} indirect jump targets. By limiting disassembly to these labeled targets, our
\emph{Time Variance Authority (\sys)} framework \textbf{both} \emph{retains
soundness} (we cannot miss valid jump targets) \textbf{and} greatly reduces
spurious code paths. The net effect is a faster, smaller rewrite that can also
enforce CET-like semantics in software. Moreover, \sys’s approach to function
pointers and external libraries (\textbf{S4}) lets developers selectively
instrument only the main binary or include libraries for a tighter security
guarantee. In either case, \texttt{endbr64} becomes a foundational element for
pruning the search space, addressing many overhead concerns that plagued
traditional superset disassembly.  

\section{System Design and Implementation}\label{sec:design}
Our work builds on Multiverse~\cite{bauman2018superset}, modernizing it for 64-bit x86 
binaries and extending its functionality with \sys. Multiverse’s core design remains sound, 
requiring only minor changes to accommodate contemporary OS and compiler behaviors. This 
section describes the overall architecture of \sys, detailing how we have leveraged 
Multiverse’s existing machinery while adding a CET-driven disassembly mode.









\subsection{Overview}

\sys divides the rewriting process into three major components—\emph{disassembly}, \emph{rewriting}, and \emph{mapping}—that work together through multiple passes to robustly transform the original code into its instrumented form.

In the \emph{disassembly} phase, \sys analyzes the binary to identify valid instructions and feasible code paths. Instead of enumerating every byte offset (as in Multiverse’s superset disassembly), \sys selectively follows legitimate indirect-branch targets by scanning for \texttt{endbr64} instructions, while still preserving all direct branch paths.

The \emph{rewriting} phase then modifies instructions so that they point to newly allocated code regions, thus adjusting both direct and indirect control flow. During this stage, \sys also inserts additional instrumentation logic—such as shadow stack checks or software-based CET-like enforcement—to strengthen security or enable profiling.

Finally, the \emph{mapping} phase maintains a lookup structure that associates original addresses with their rewritten counterparts. At runtime, this mapping is consulted whenever execution might revert to code segments whose execute permissions have been revoked, ensuring that the flow is correctly redirected into the instrumented binary.

Like Multiverse, \sys performs these steps in two passes. The first pass scans and collects all offsets marked for relocation or patching and builds the mapping table. The second pass uses that mapping to produce the final binary, ensuring that all relative and absolute addresses remain accurate.

\subsection{Disassembly}
Multiverse guarantees a sound disassembly of the original binary by iterating over every 
instruction offset. While this \emph{superset disassembly} ensures no code path is 
missed, it can excessively inflate the number of instructions to be examined and mapped. 
On modern binaries, this can lead to significantly longer analysis and instrumentation 
times.

\subsubsection{TVA Disassembly}
\sys addresses the overhead of superset disassembly by leveraging CET instructions. 
CET dictates that indirect jumps must target \texttt{endbr64}, effectively labeling 
legal landing pads within a function. Whenever \sys encounters a direct branch, it 
follows it normally. However, for indirect branches, \sys only initiates new disassembly 
paths at \texttt{endbr64} offsets. This strategy reduces the number of spurious offsets 
that might never be legitimately branched to, thereby cutting down the total disassembly 
work.

From a security standpoint, \sys also enforces CET-like guarantees in software. If an 
indirect branch attempts to jump to an offset that lacks a preceding \texttt{endbr64}, 
execution will be halted or trapped, mirroring the hardware behavior on CET-capable CPUs.

\subsection{Mapping}
Like Multiverse, \sys adopts a two-pass mapping strategy. In the first pass, all potential 
jump targets are collected: direct branches, \texttt{endbr64} instructions, and (for call/return 
semantics) the instructions immediately following a call. The second pass synthesizes the 
new code sections and writes out the final instrumented binary. Any relocated instruction 
or data reference is updated to point to the new addresses according to the collected 
mapping. 

\subsubsection{\sys Mapping}
In practice, \sys expands on Multiverse’s mapping logic by restricting indirect branch 
targets to \texttt{endbr64}. This ensures every valid dynamic target is tracked. Attempts 
to jump to unmapped addresses in the original code region trigger a segfault, which our 
runtime mechanism intercepts and safely redirects or halts. Thus, even CPUs without CET 
extensions gain a CET-like level of control-flow validation.

\subsection{Executing Lookups}\label{sec:implementation}
A key operational difference in \sys is how it prevents execution in the original code 
region. Specifically, \sys revokes execute permissions from those pages, so any attempt 
to run code there results in a segfault. This ensures that the new, instrumented code 
is always the point of control flow, even for external libraries returning into what 
would normally be the original text section.

\subsubsection{Handling Segfaults}
For uninstrumented libraries, such as \texttt{libc}, the system initially follows 
Multiverse’s design of attempting to map the call address, then the return address if 
necessary. In some scenarios, the original address might get used on return, which 
previously risked resuming execution in the old code. With execute permissions removed, 
such a return now triggers a segfault. Our segfault handler can then capture the 
offending address and consult the mapping table to find the correct instrumented address. 
Execution resumes safely in the rewritten code, maintaining consistent control-flow 
tracking and preventing any reversion to the original, uninstrumented regions.

This mechanism allows \sys to function seamlessly alongside legacy binaries and system 
libraries, bridging gaps where full-library instrumentation might not be feasible or 
desirable. By relying on the segfault-based redirection, we minimize risk while still 
maintaining robust instrumentation for the main application code.

Rather than proactively remapping the return address (as Multiverse originally did), we now 
allow it to remain unchanged. Upon return, the process will segfault if it attempts to use 
the original address, at which point we can remap it on demand. This preserves the original 
return address for any external library logic that relies on it, while still ensuring that 
execution ultimately transitions back into our instrumented code.

\paragraph{\sys{} return addresses}
Because we rely on segfaults to remap control flow from returning from functions we need those to be valid map locations.
This means that we, unfortunately, need to relax valid mapping from just the \texttt{endbr64} instructions that a normal CET processor would allow and include instructions directly after a call instruction as well.
However, with a better implementation of a shadow stack we can also mitigate this issue as well.

\subsubsection{Mapping Functions}
The address resolution itself is done in two stages.
First a local mapping is used to check if, and translate if it is, an address is in the original code area of the current binary.
If it is then we use Multiverse's constant time lookup to determine the new offset.
If the offset is \texttt{0xffffffff} then we deem it as illegal and exit.

Otherwise, the address will be outside the current mapped region and we need to consult the global mapping in order to determine if the correct mapping function to use, and tail call into it.

The implementation of the mapping functions has deviated slightly to conform to generated code becoming position independent by nature of the widespread introduction of ASLR.
Instead of having special cases for PIC, the default is to use \texttt{RIP} relative offsets to get the correct base address to determine code regions.

\subsubsection{Executing Lookups} \label{sec:Lookups}
As we discussed in the last section, with a mapping in place we need some way to handle the actual translation of original addresses into new addresses.
This is the job of the lookup function.
As discussed in the section on fixing Multiverse, we removed execute permissions from the original code pages.
This mean that if we try to execute in the original code section, the program will now segfault.
In case that was not the intended behaviour, we add a segfault handler which calls the mapping function to get us back to the new code section to allow the program to finish running.

The process for looking up the new address for an original address is done in up to three stages.
First a local lookup is done, if it succeeds then we return immediately.
Otherwise the local lookup defers to a global lookup, which checks for the appropriate local lookup function to call and passes it along.
Finally it reaches the proper local lookup function which will handle the translation before returning.

This overall process is the same between Multiverse and \sys.

\paragraph{Local Lookup}

\begin{figure}
    \centering
\begin{lstlisting}
    lookup:
        push rbx
        mov rbx,rax
       lea rax, [rip + lookup - {lookup_offset}]
        add rbx, {self.context.newbase - base}
        sub rbx, rax
        jb outside
        cmp rbx, {size}
        jae outside
        mov ebx,[rax+rbx*4+{mapping_offset}]
        cmp ebx, 0xffffffff
        je failure
        add rax,rbx
        pop rbx
        ret
    outside:
        add rbx, rax
        sub rbx, {self.context.newbase - base}
        mov rax,rbx
    mov rbx, [0x56780008]
    xchg rbx, [rsp]
    add rsp, 8
    jmp [rsp-8]
    failure:
        hlt
\end{lstlisting}
    \caption{Local lookup function}
    \label{fig:local-lookup}
\end{figure}

In Figure \ref{fig:local-lookup} we see the assembly for executing 
First it calculates the base address of where the program is loaded.
The it gets the offset from the base that the address to be translated is at.
If that address is outside the range, either it is at a negative offset or it is at an offset greater than the original size of the program, then we defer to the global lookup.
Otherwise we use our mapping as a lookup table in order to do a constant time translation.
If the offset is \texttt{0xffffffff} then we deem it as illegal and halt the program.
Otherwise we add it to the base address before returning it.

The implementation of the mapping functions has changed slightly from Multiverse's original form because most code is position independent code (PIC) by nature of the widespread introduction of ASLR.
Instead of having special cases for PIC, the default is to use the offset from the base of where the program was loaded.
This vastly reduces the complexity of having to handle different types of instrumentation differently.

\paragraph{Global Lookup}
\begin{figure}
    \centering
\begin{lstlisting}
    glookup:
        ...
        mov rcx, 0x56780000
        mov rbx, [rcx]
        xor rdx, rdx
    searchloop:
        cmp rbx, rdx
        je failure
        add rcx, 24
        mov r10, [rcx+8]
        neg r10
        add r10, rax
        cmp r10, [rcx+16]
        jle success
        inc rdx
        jmp searchloop
    success:
        ...
    external:
        pop r10
        pop rdx
        pop rbx
        pop rcx
        test rbx, rbx
        jz skip
        mov [rsp-64],rax
        mov rax,[rsp+8]
        call glookup
        mov [rsp+8],rax
        mov rax,[rsp-64]
        skip:
        ret
    failure:
        hlt
\end{lstlisting}
    \caption{Global lookup function}
    \label{fig:global-lookup}
\end{figure}

In Figure \ref{fig:global-lookup} we see the assembly for the global lookup.
The preamble for the global lookup includes storing a whole slew of registers that we'll need.
We have to do this because there's no concept of caller saved registers because we don't do analysis on what registers the program actually uses.
Then we have a loop for searching for the proper mapping region.
When we find the region that corresponds to the address we're mapping, we can call the corresponding lookup function, here stored in \texttt{rcx} on line 21.
If the look up function is not null then we use the mapped value from the lookup function.
If the lookup function is null, then there is no lookup function and therefore it's an external lib.
In the case of external libs we have to determine if we need to remap the return address, which is at a known offset on the stack.
This determination is passed in on the \texttt{rbx} register and we'll see it when we look at how we rewrite call and return instructions.

\subsection{Rewriting}
The tactics by which we rewrite control flow remain the same as Multiverse.

For direct control flow, once we have a mapping we can immediately replace the offsets with the correct mapping.
Branches are padded to be 4 bytes in case they were assembled to only have a 1 byte immediate value, so that we stay in sync with the mapping.

Indirect branches are rewritten to use the mapping functions in order to translate from the old address to the new address.

\begin{lstlisting}
    mov [rsp-64], rax
    mov [rsp-72], rbx
    mov rax, target
    mov rbx, isjmp
    call lookup
    mov [rsp-8], rax
    mov rax, [rsp-64]
    mov rbx, [rsp-72]
    jmp [rsp-8]
\end{lstlisting}

Here we match Multiverse's method with an added \texttt{rbx} store to use it to pass in whether the instruction is a call or not.
In this case we will indicate that it is an indirect jump so that the mapping function will not try to remap the return address if we are going into an external address.

\subsubsection{Calls}
On the other hand an indirect call will have a slightly longer variant to determine the return address is calculated with a \texttt{lea}.
It is done this way because we are now dealing with PIC, so we cannot know the correct value for the return address at instrumentation time as we do not know where it will be loaded.

\begin{lstlisting}
    mov [rsp-64], rax
    mov [rsp-72], rbx
    mov rax, target
    mov rbx, isjmp
    push rbx
    lea rbx, return_address
    xchg rbx,[rsp]
    call lookup
    mov [rsp-8], rax
    mov rax, [rsp-56]
    mov rbx, [rsp-64]
    jmp [rsp-8]
\end{lstlisting}

We restore \texttt{rax} from \texttt{rsp-56} instead of \texttt{rsp-64} because we pushed the return address.
And similarly for \texttt{rbx}.

\subsubsection{Return}
Because we push the original code address for the return address, we need to remap that before we return.

\begin{lstlisting}
    mov [rsp-56], rax
    pop rax
    push rbx
    mov rbx, 0
    call $+%s
    pop rbx
    mov [rsp-8], rax
    mov rax, [rsp-56]
    jmp [rsp-8]
\end{lstlisting}

Here we store away \texttt{rax} before grabbing the return address and setting \texttt{rbx} to remind the mapping functions that we are not a call, so that it does not try to remap something as the return address when it determines that we are going to an external address.
Once we get the mapped return address we just need to restore our registers and then we can jump.

Together, these changes allow \sys{} to manage the intricacies of 64-bit binary 
analysis and rewriting while preserving compatibility with modern Linux environments. 
By adapting the disassembly engine, addressing modes, and calling conventions, \sys{} 
can confidently handle large, complex binaries without losing soundness or precision.
\subsection{CET}
For CET-enabled binaries Multiverse can, in theory, support, without modifications to the framework, an instrumentation pass that gives the resulting binary CET enforcement in software, if you have a non-supporting CPU.
The insight of \sys{} is to use CET instructions to augment the disassembly process.
Since we know that the execution expects to use the enforced semantics of CET instructions we can limit the disassembly to only start from CET indirect jump target instruction, \texttt{endbr64}.

\subsection{Address Map Size}
The other change is for execution, since we know the only places we'll be doing an indirect jump to is an indirect jump target, that's the only things that we'll need to add to our mapping are the indirect jump targets.
Direct jump targets don't need to be mapped because the instrumentation step puts in the offsets for those, so we can omit the mappings for those.
The end result is we reduce the disassembled instruction count while improving the security properties of the final binary.

\section{Use Cases}

Our framework enables high-value instrumentation and security features in
settings that typically require specialized hardware or kernel-level support.
In this section, we illustrate two representative applications: lightweight
tracing and analysis\cite{wang2022memliner}, and software-based enforcement of CET-like semantics\cite{tymburiba2019multilayer}.
By instrumenting binaries either offline or at runtime, developers can
achieve lower overhead than traditional kernel-level tracing methods, while
still benefiting from control-flow protections that typically require modern
CPU features. In legacy or virtualized environments lacking dedicated CET
hardware, our approach provides a transparent way to retrofit security and
observability into existing applications.

\subsection{Lightweight Tracing and Analysis}
\label{sec:lightweight-tracing}

Profiling or debugging complex software often relies on tracing frameworks
that intercept function boundaries or specific code locations at runtime.
While these approaches (e.g., \texttt{perf}, \texttt{uprobe}\cite{bpftime}) excel at flexibility,
they can suffer from high overhead due to breakpoints, context switching,
and frequent kernel intervention—particularly detrimental in real-time or
high-throughput systems. By contrast, our framework reduces or eliminates
these sources of overhead by injecting analysis code directly into the
binary. Whether performed offline (static rewriting) or applied dynamically,
this embedding avoids kernel-level traps and consolidates instrumentation
logic into a single, self-contained executable. As a result, developers can
achieve lower-latency tracing that remains easy to deploy: there is no need
for specialized kernel modules or repeated probe registration. The final
executable transparently includes all necessary trace hooks, allowing for
predictable performance even under demanding workloads.

\subsection{Software CET Emulation}
\label{sec:software-cet}

Although recent hardware features (e.g., Intel CET or AMD’s Shadow Stack) offer 
built-in safeguards against control-flow attacks such as Return-Oriented Programming (ROP) 
and Jump/Call-Oriented Programming (JOP), many deployed systems lack these capabilities. 
Rather than requiring specific processors, our approach enforces CET-like semantics 
entirely in software, providing two main defenses for legacy binaries and virtualized 
environments.

\textbf{shadow stack protection} preserves return-address integrity by instrumenting 
every \texttt{call} and \texttt{return} instruction. When a function call is invoked, 
the return address is pushed both onto the standard stack and onto a separate, 
software-maintained shadow stack. At \texttt{return}, these two addresses are compared; 
execution continues only if they match. If not, the process is halted, thwarting 
ROP-style attempts to corrupt the return address. While hardware shadow stacks on 
newer CPUs perform a similar check in a protected memory region, our software approach 
makes this defense available on any commodity x86\_64 processor.

\textbf{indirect branch validation} emulates the CET requirement that 
indirect jumps or calls must land on a specially labeled target (\texttt{endbr64}). 
Upon any indirect control transfer, our system checks whether the destination 
offset corresponds to a valid \texttt{endbr64} site in the instrumented code. 
If it does not, execution is blocked. This design ensures that attackers cannot 
redirect control flow into mid-instruction bytes or other unintended regions, 
mirroring the hardware-enforced checks on CET-enabled CPUs.

Together, these two protections address the main pillars of CET without relying 
on specialized hardware. By rewriting a binary to insert \texttt{endbr64} markers 
and shadow stack instrumentation, our framework can detect both invalid return 
addresses and illegal branch targets. In practice, the added runtime overhead 
is modest, and developers can seamlessly retrofit these defenses into existing 
binaries or libraries. This software-based CET emulation thus offers broad 
compatibility and a compelling security advantage, especially in legacy or 
virtualized deployments where hardware CET is not yet available.

\subsection{Replacing DBI}
\label{sec:killer-app}

\begin{figure}
    \centering
    \includegraphics[width=0.4\textwidth]{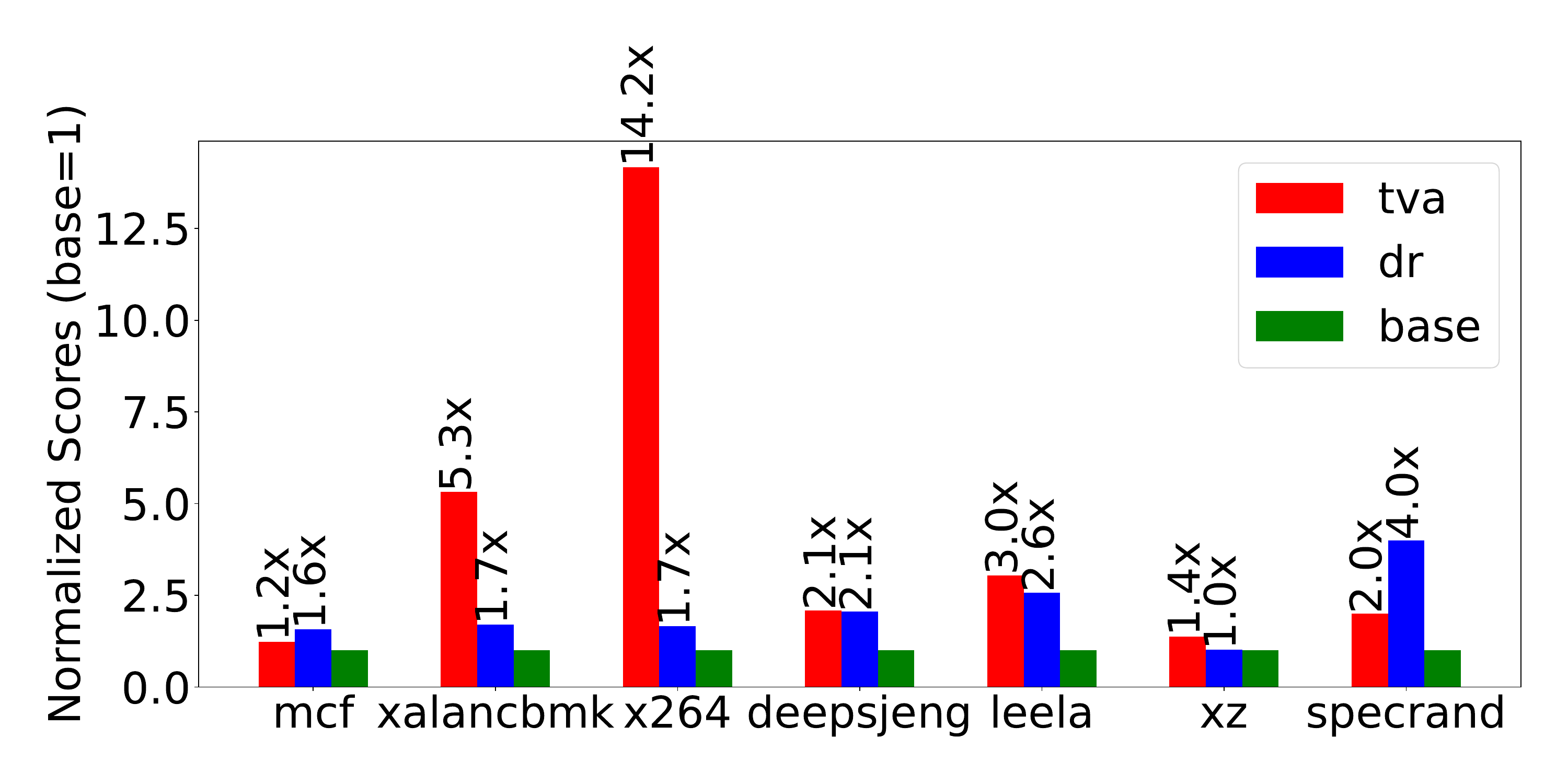}
    \caption{Runtime comparison for \sys, Null tool DynamoRIO, and uninstrumented on CPU2017}
    \label{fig:null-tool-comparison}
\end{figure}

Dynamic binary instrumentation (DBI) frameworks\cite{villa2019nvbit,guo2015accelerating} such as DynamoRIO~\cite{dynamorio} 
provide a powerful mechanism to insert analysis or security checks \emph{during} runtime. 
They work by intercepting execution and translating basic blocks on the fly, typically 
caching instrumented code fragments in a code cache. While flexible, DBI approaches 
carry several drawbacks that \sys (and static rewriting in general) can overcome:

DBI frameworks must maintain a code cache and translation engine, leading to significant overhead in CPU- and memory-intensive workloads. In contrast, \sys’s static rewriting integrates instrumentation directly into the binary, eliminating the need for repeated runtime translations. As our evaluation shows, this reduces runtime slowdown and is crucial for latency-sensitive or large-scale workloads such as HPC or high-throughput servers. While there is some slowdown in \Cref{fig:null-tool-comparison}, we achieve a 10x speedup in memtrace. Compared to the null tool performance, we find that the instrumentation overhead and calling into external C/C++ code is very expensive \cite{expensive}.

Because DBI tools modify code streams at runtime, they can introduce jitter, causing variable execution times as new paths are discovered and translated. For real-time or time-critical applications, this unpredictability can be unacceptable. By applying instrumentation \emph{offline}, \sys ensures that the resulting executable maintains uniform behavior with stable timing characteristics.

Dynamically instrumented programs often rely on hooking into system loaders or attaching to processes via debuggers, requiring specific runtime privileges such as \texttt{LD\_PRELOAD} usage or container permissions that may be restricted in production or sandboxed environments. In contrast, \sys produces a fully self-contained \emph{rewritten} binary that can be deployed like a normal executable, with no special privileges or environment variables required.

DBI frameworks also impose a substantial memory overhead to maintain the code cache, analysis data structures, and runtime instrumentation logic. By adopting a static rewriting approach, \sys achieves a memory footprint comparable to the original binary, with only a minimal overhead for instrumentation stubs. This efficiency is particularly valuable in resource-constrained environments or large-scale servers, where additional memory usage is costly.

\section{Evaluation}\label{sec:evaluation}
This section presents our evaluation of the \sys prototype.  We implement the \numcases{} case studies on \sys.  We answer the following questions:

\begin{smitemize}
\item How does \sys's performance overhead compare to Multiverse? 
\item Why does \sys impose lower runtime overhead than existing tools, \ie from where does the system's performance advantage arise?
\item How \sys can be useful in rewriting that prior work is not working? What's the killer application for \sys?
\end{smitemize}

\subsection{Experimental Setup}

We evaluate \sys running on a dual-socket AMD EPYC 7452 (32 cores, 64 SMT 2.35 GHz, 128 MB LLC) with 256 GB DDR4 memory.
Unless otherwise mentioned, each number listed below is the average of 10 trials.
We report averages using geometric mean when that is appropriate.

\begin{figure}
        \centering
    \includegraphics[width=0.4\textwidth]{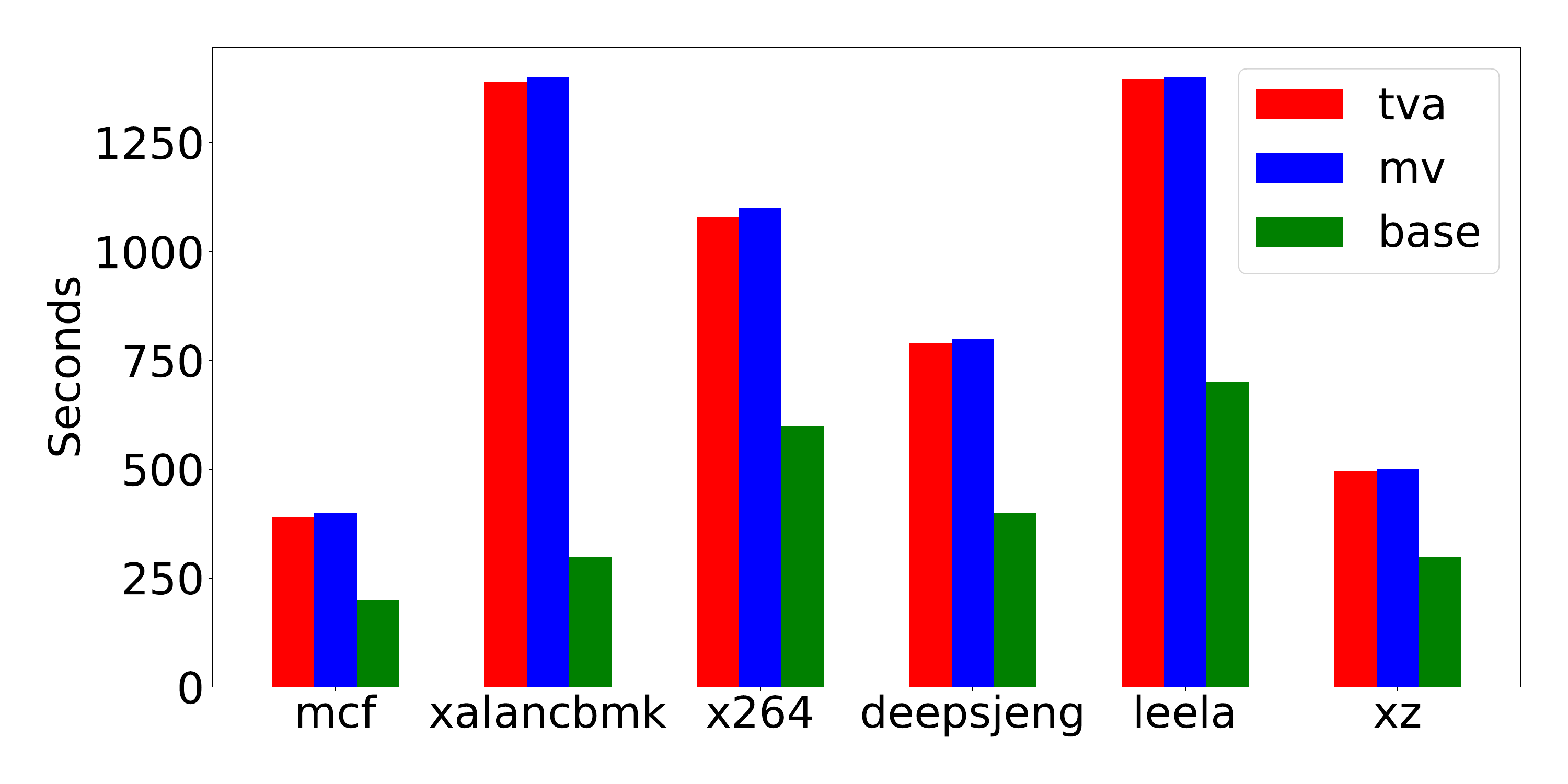}
    \caption{Runtime comparison for \sys, MV, and uninstrumented on CPU2017}
    \label{fig:perf-comparison}

\end{figure}
\vspace{-0.3cm}
\begin{figure}

    \centering
    \includegraphics[width=0.4\textwidth]{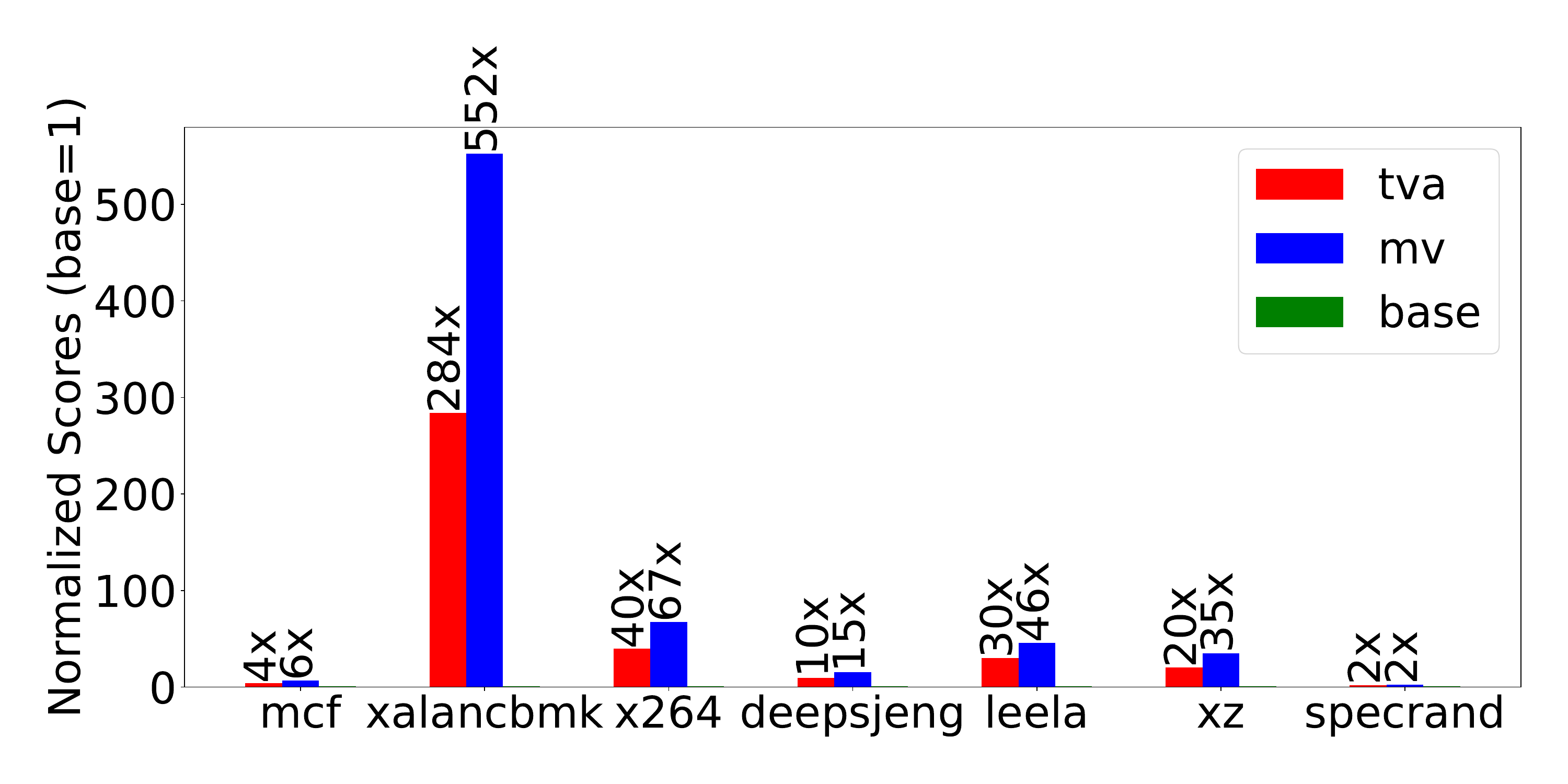}
    \caption{Compile time comparison for \sys, MV, and uninstrumented on CPU2017}
    \label{fig:compile-comparison}
\end{figure}
\begin{figure}
    \centering
    \includegraphics[width=0.4\textwidth]{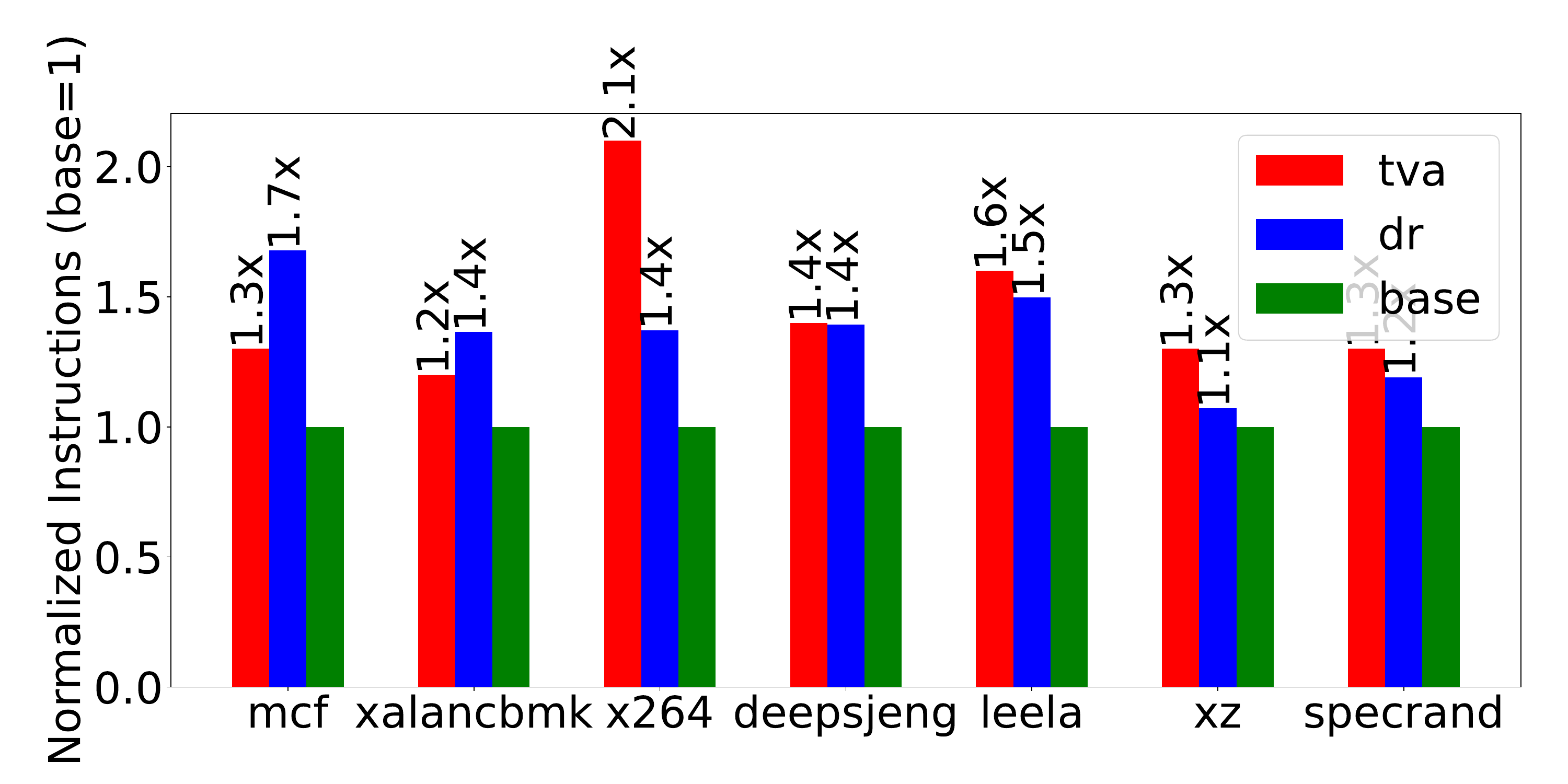}
    \caption{Instruction Count comparison for \sys, Null tool DynamoRIO, and uninstrumented on CPU2017}
    \label{fig:null-tool-instruction-comparison}
\end{figure}
\begin{figure}
    \centering
    \includegraphics[width=0.4\textwidth]{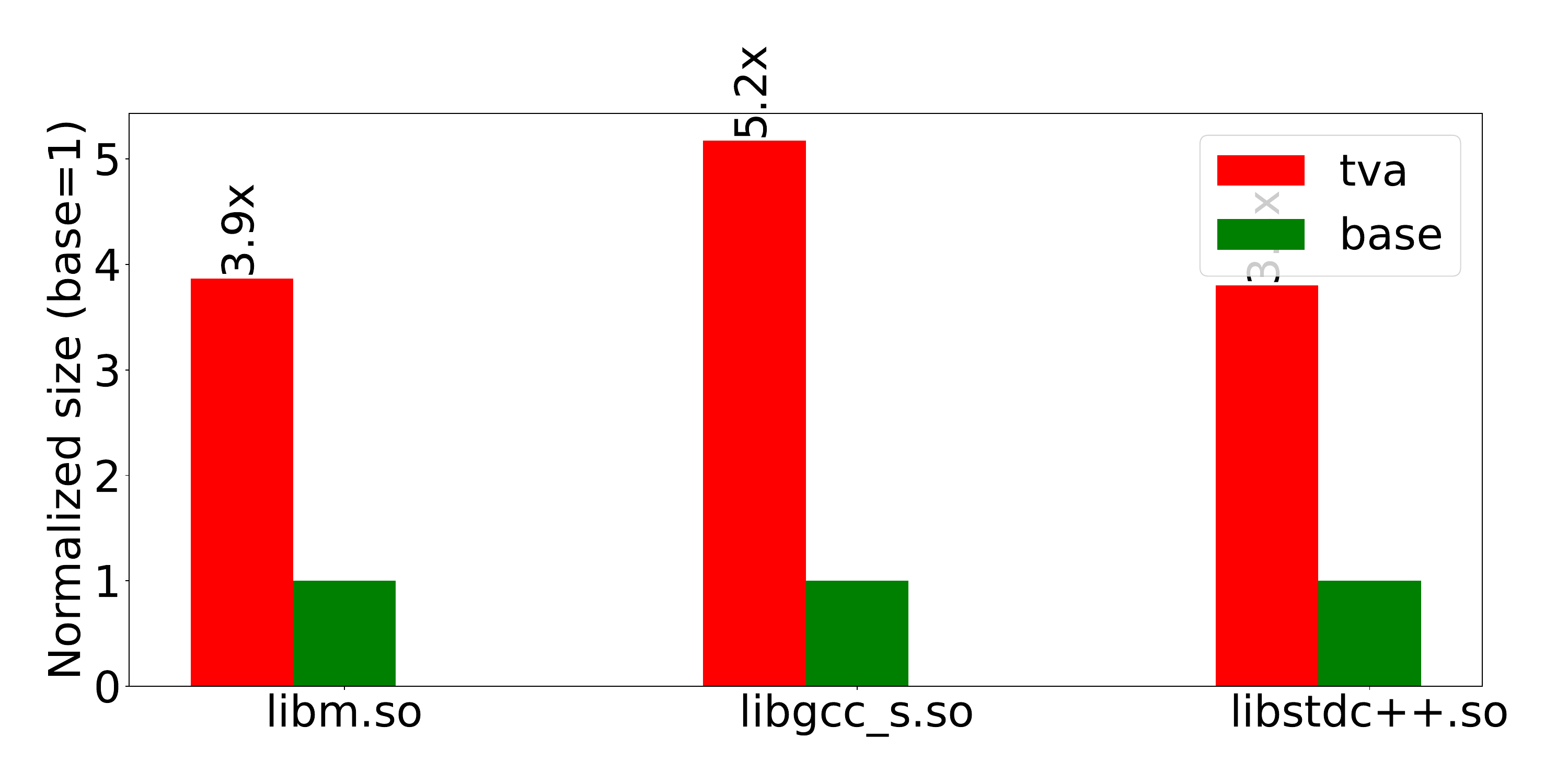}
    \caption{Library size comparison for \sys}
    \label{fig:lib-size}
\end{figure}
\subsection{Benchmarks}
While our evaluation centers on SPEC2017, there are certain workloads we cannot handle due to assumptions inherent in our approach. Specifically, our tool does not support self-modifying executables, ruling out the \texttt{gcc} benchmark in SPEC2017, which changes its own code at runtime. Other workloads also pose unique challenges: \texttt{perlbench} triggers failures when destructors attempt to call into code sections not recognized by our rewriter, \texttt{omnetpp} references addresses in \texttt{libgcc} that remain unmapped after rewriting, and \texttt{leela} implicitly depends on \texttt{libm} in ways that complicate static analysis. These issues highlight a broader limitation of static instrumentation when confronted with dynamic or low-level runtime behaviors. 
\subsection{Instrumentation}

In figure \ref{fig:compile-comparison} we compare the usage of Multiverse and \sys{}.
While we can disassemble and instrument any binary with multiverse and any binary with CET-enabled with TVA, we can only instrument the original binary.
If the binary is self-modifying, where it changes code in the original code region, then the instrumented execution will be different from the original as the instrumented execution will not reflect any of those modifications.
If the program relies on JIT and it stores the resulting instructions outside the original code area, that will be executed but not instrumented at all.

The instrumentation time metric was measured by running the two systems with a null instrumentor, where there is no added instrumentation.
We are able to see 30\% speedups with \sys{} over Multiverse.
Instrumentation time with the null instrumentor is linear with number of instructions that need to be disassembled and reassembled.
Therefore we see that \sys is able to prune a third of the potential instructions that Multiverse otherwise tries to disassemble.

\subsection{Performance Overhead}
\label{sec:perf-overhead} 

\begin{table*}[]
    \centering
    \begin{tabular}{|rr|r|r|r|r|} \hline
        \multirow{2}{*}{Binary} &\multirow{2}{*}{Instrumenter} &\multicolumn{2}{|c|}{Original} & \multicolumn{2}{|c|}{New}\\
        \cline{3-6}
        && Text Size& File Size& Text Size& File Size\\
         \hline
\multirow{2}{*}{mcf}     &TVA& 19673   &  125128   &  106001  &  280800 \\
&MV& 19673 & 125112&  194775& 369568\\ \hline
\multirow{2}{*}{xalancbmk}&TVA&4711945  &79367160  &26689195  &106105328\\
&MV&4708409&79362016&46684765&126096800\\ \hline
\multirow{2}{*}{x264}     &TVA& 560673  & 3330528  & 2910577  & 6292608\\
&MV& 560401& 3330416& 5333702 & 8715728\\ \hline
\multirow{2}{*}{deepsjeng}&TVA&  82697  &  451520  &  432902  &  935376 \\
&MV&  82585&  451488&  795836& 1298304\\ \hline
\multirow{2}{*}{leela}    &TVA& 218421  & 4596272  & 1165596  & 5813344\\
&MV& 218421& 4596256& 2068784 & 6716528\\ \hline
\multirow{2}{*}{xz}       &TVA& 144197 &  992824 &  768822 & 1812032 \\
&MV& 143541 &  992560 & 1379073 & 2422288\\
         \hline
    \end{tabular}
    \caption{Size before and after instrumentation}
    \label{tab:instr-size}
\end{table*}

Looking at Table \ref{tab:instr-size} we can see how \sys and Multiverse compare on increasing memory footprint.
We see that \sys uses almost half the new text size as Multiverse does.
That is the only place that we currently see savings, though it's possible that future work may be able to find more space savings by condensing the mapping used to translate addresses.

\begin{table*}[]
    \centering
    \begin{tabular}{|rr|r|r|r|r|r|r|r|} \hline
        \multirow{2}{*}{Binary} &\multirow{2}{*}{Instrumenter} &\multirow{2}{*}{Instructions}&\multicolumn{2}{|c|}{Direct} & \multicolumn{2}{|c|}{Indirect}&Conditional\\
        \cline{4-7}
        &&  & Call & Jump & Call & Jump & Jump\\
         \hline
\multirow{2}{*}{mcf}     &TVA& 9584   &  304   &  276  &  62&  42&  900 \\
&MV& 36428 & 372&  404& 160&  120&  1884\\ \hline
\multirow{2}{*}{xalancbmk}&TVA&2267916  &139268  &101438  &59606&  4280&  165850\\
&MV&8662788&159688&120076&71640&  27782&  435402\\ \hline
\multirow{2}{*}{x264}     &TVA& 262466  & 6236  & 6166  & 1136&  172&  16330\\
&MV& 1044136& 9628& 9234 & 2218&  1878&  43538\\ \hline
\multirow{2}{*}{deepsjeng}&TVA&  38492  &  1620  &  1436  &  4 &  68&  4192\\
&MV&  151670&  2110&  1852& 110&  548&  7274\\ \hline
\multirow{2}{*}{leela}    &TVA& 96010  & 7778  & 3862  & 8&  156&  8380\\
&MV& 389932& 9266& 4970 & 378&  1222&  20126\\ \hline
\multirow{2}{*}{xz}       &TVA& 70574 &  3010 &  2540 & 150 &  126&  6012\\
&MV& 264152 &  4094 & 3380 & 328&  806&  13378\\
         \hline
    \end{tabular}
    \caption{Instructions Instrumented}
    \label{tab:instr-count}
\end{table*}

With Table \ref{tab:instr-count} we see the counts of various types of instructions that are being instrumented.
Across the board, we see huge decreases in total instructions disassembled as well as in very expensive categories, like indirect calls and jumps.

\Cref{fig:perf-comparison} shows the execution times of the SPEC CPU2017 
benchmarks under three different configurations: our proposed tool (\texttt{tva}), 
the original Multiverse (\texttt{mv}), and an uninstrumented baseline (\texttt{base}). 
These benchmarks include 505.mcf\_r4, 523.xalancbmk\_r4, 525.x264\_r4, 531.deepsjeng\_r4, 541.leela\_r4, and 557.xz\_r4. 

For each benchmark, the \texttt{tva} (red bars) and \texttt{mv} (blue bars) bars 
measure the runtime of the statically-instrumented binaries, while the \texttt{base} 
(green bars) corresponds to the uninstrumented executable. As expected, both 
instrumented configurations incur an overhead above the baseline. However, \texttt{tva} 
consistently exhibits lower or comparable overhead when compared to \texttt{mv}. 

    In benchmarks like 505.mcf\_r4 and 531.deepsjeng\_r4, we observe a moderate increase 
in runtime over \texttt{base}, largely due to additional control-flow checks. On other, 
more complex benchmarks such as 523.xalancbmk\_r4 and 525.x264\_r4, the overhead 
becomes more pronounced, yet \texttt{tva}'s runtime still slightly outperforms that 
of \texttt{mv}. This improvement arises because \texttt{tva} avoids over-disassembly 
by leveraging \texttt{endbr64} instructions, thus reducing the amount of code 
that needs to be instrumented. 


\subsection{Performance Optimization}
Below, we highlight two main considerations: (1) whether to instrument external libraries, 
and (2) avoiding unnecessary address translations at runtime.

\paragraph{Instrumenting vs.\ Ignoring Shared Libraries.}
A fundamental trade-off in \sys{} is whether to apply rewriting to shared libraries 
 in addition to the main executable. 
Instrumenting these libraries can improve security coverage (and, in some instrumentation scenarios, 
increase observability), because every indirect call within library code also becomes constrained 
by CET-like checks. However, this may significantly increase overall rewriting time (especially for 
large libraries) and inflate the size of the final instrumented binaries.

By contrast, ignoring libraries---rewriting only the main executable---leads to \emph{faster} 
instrumentation (often by up to 30--40\% in our experiments) and keeps final binary sizes closer 
to the uninstrumented baseline. The trade-off is that any indirect calls within library code remain 
uninstrumented. As a safeguard, \sys{} still revokes \emph{execute} permissions for the original 
(baseline) code pages of the main binary, so returns from library code into old addresses trigger 
a segfault (see Section~\ref{sec:implementation}). At that point, the segfault handler looks up 
the correct rewritten address and resumes execution in the instrumented code. In practice, we find 
that this “on-demand” mapping does not add noticeable runtime overhead for most SPEC~CPU2017 
benchmarks, as shown in Figure~\ref{fig:perf-comparison}.

\paragraph{Segfault-Based Boundary Handling.}
When \sys{} does \emph{not} instrument libraries, any library-to-executable returns (or callbacks) 
may still contain the original return address in the call stack. Because \sys{} marks 
the original code pages as non-executable, attempts to jump there cause a segfault. Our segfault 
handler consults \sys{}’s mapping to translate the old address into the new, instrumented offset. 
We found that for typical applications, these segfaults are rare enough in practice that the overall 
performance impact is minimal. This approach also simplifies deployment by avoiding the need to 
statically link large libraries or create fully instrumented versions of them.
\vspace{-0.3cm}

\paragraph{Minimizing Address-Translation Calls.}
Another optimization is to reduce the number of address translations (\ie calls to 
\texttt{lookup()} or similar stubs) at runtime. \sys{} already prunes indirect jumps that 
cannot legally occur (due to CET’s \texttt{endbr64} markers), but it can further minimize 
translation overhead by caching translation results for frequently used function pointers. 
If the binary makes heavy use of function pointer arrays or virtual calls (as in \texttt{C++}), 
memoizing their rewritten destinations can reduce repeated lookups. Our prototype implementation 
currently uses a simple hash-based cache; in future work, more sophisticated caching heuristics 
(\eg hardware-assisted or kernel-based caching) could further improve performance.

\paragraph{Impact on Instrumentation Time and Binary Size.}
Ignoring libraries yields a 1.2$\times$--1.3$\times$ 
faster instrumentation phase on average, but leads to a 10--15\% chance of 
on-demand segfaults whenever external code returns to the original code region. 
For workloads that repeatedly cross from the main binary into library functions (and back), 
instrumenting libraries can amortize segfault overhead at runtime. However, if those library 
functions are large and seldom rely on indirect calls, rewriting them can inflate the final 
binary size by 1.5$\times$--2$\times$ (see Figure~\ref{fig:lib-size}).
\section{Related Work}
\label{sec:related}

Binary reassembly and rewriting have received considerable attention from both academia and industry.
Early techniques employed patch-based modifications, wherein instrumentation code was inserted directly into original binary locations.
However, such in-place patching imposes severe constraints on available space and can unintentionally overwrite embedded data.
Subsequent approaches introduced table-driven solutions or symbolization to achieve a more robust and flexible rewriting \cite{flores2020datalog, dinesh2020retrowrite, williams2020egalito}.
Despite these advances, disassembly and pointer identification inaccuracies remain significant challenges in the presence of highly optimized or obfuscated x86-64 binaries.

To address these limitations, Kim~\emph{et~al.}~\cite{kim2025towards} propose \textbf{SURI}, a framework for sound reassembly of modern x86-64 binaries.
Unlike prior solutions that often rely on ad-hoc disassembly heuristics, SURI constructs \emph{superset CFGs}:
it recursively follows direct branches and statically analyzes jump tables to over-approximate potential indirect branch targets.
SURI also performs backward slicing for each indirect branch (e.g., \texttt{jmp rdx}) to discover symbolic expressions indicating possible jump-table addresses.
By continuously applying dataflow analysis whenever new indirect edges appear, SURI ensures that no valid control-flow paths are missed.
As a result, SURI significantly improves the soundness of reassembly, particularly for position-independent and Control-Flow Enforcement Technology (CET)-enabled binaries \cite{kim2025towards}. However, they are not sound since they do not make the target binary think they are not instrumented.

\vspace{-0.3cm}

\section{Conclusion}
\label{sec:conclusion}

We presented a 64-bit extension of Multiverse that preserves the strengths of
superset disassembly while integrating with Intel CET to limit indirect jumps
to \texttt{endbr64} targets. By leveraging hardware labels (or their software
equivalents for non-CET CPUs), our framework eliminates many spurious paths in
the rewriting process, reducing instrumentation overhead and maintaining strong
control-flow integrity guarantees. Across SPEC~CPU2017 and real-world applications, we demonstrated up to a
1.3$\times$ improvement in instrumentation speed, a 2\% runtime overhead
reduction. These results highlight the power of CET-guided analysis
for efficient, sound binary rewriting. Future work may extend this methodology
to other ISAs or further explore how CET-like annotations can streamline
additional security and performance optimizations.

\bibliographystyle{ACM-Reference-Format}
\bibliography{uctest}

\end{document}